\def\ltilde#1{\mathord{\mathop{\kern 0pt #1}\limits_\sim}}
\typein[\sorb]{Enter ``s'' (for small) or ``b'' (for big)}
\newlength{\absize}

\if s\sorb \documentstyle{article}
\renewcommand{\baselinestretch}{1.5}
\renewcommand{\arraystretch}{1.5}
\begin{document}
\date{}
\pagestyle{empty}
\thispagestyle{empty}
\renewcommand{\thefootnote}{\fnsymbol{footnote}}
\newcommand{\starttext}{\newpage\normalsize
\pagestyle{plain}
\setlength{\baselineskip}{4ex}\par
\twocolumn\setcounter{footnote}{0}
\renewcommand{\thefootnote}{\arabic{footnote}}
}
\else
\documentstyle[12pt]{article}
\setlength{\absize}{6in}
\setlength{\topmargin}{-.5in}
\setlength{\oddsidemargin}{-.3in}
\setlength{\evensidemargin}{-.3in}
\setlength{\textheight}{9in}
\setlength{\textwidth}{7in}
\renewcommand{\baselinestretch}{1.5}
\renewcommand{\arraystretch}{1.5}
\setlength{\footnotesep}{\baselinestretch\baselineskip}
\begin{document}
\thispagestyle{empty}
\pagestyle{empty}
\renewcommand{\thefootnote}{\fnsymbol{footnote}}
\newcommand{\starttext}{\newpage\normalsize
\pagestyle{plain}
\setlength{\baselineskip}{4ex}\par
\setcounter{footnote}{0}
\renewcommand{\thefootnote}{\arabic{footnote}}
}
\fi

\newcommand{\preprint}[1]{\begin{flushright}
\setlength{\baselineskip}{3ex}#1\end{flushright}}
\renewcommand{\title}[1]{\begin{center}\LARGE
#1\end{center}\par}
\renewcommand{\author}[1]{\vspace{2ex}{\Large\begin{center}
\setlength{\baselineskip}{3ex}#1\par\end{center}}}
\renewcommand{\thanks}[1]{\footnote{#1}}
\renewcommand{\abstract}[1]{\vspace{2ex}\normalsize\begin{center}
\centerline{\bf Abstract}\par\vspace{2ex}\parbox{\absize}{#1
\setlength{\baselineskip}{2.5ex}\par}
\end{center}}

\newcommand{\ds}{\displaystyle}
\newcommand{\tr}{\,{\rm tr}}
\newcommand{\be}{\begin{equation}}
\newcommand{\ee}{\end{equation}}
\def\cL{{\cal L}}
\def\cO{{\cal O}}
\newcommand\etal{{\it et al.}}
\newcommand{\prl}[3]{Phys. Rev. Letters {\bf #1} (#2) #3}
\newcommand{\prd}[3]{Phys. Rev. {\bf D#1} (#2) #3}
\newcommand{\npb}[3]{Nucl. Phys. {\bf B#1} (#2) #3}
\newcommand{\plb}[3]{Phys. Lett. {\bf #1B} (#2) #3}
\newcommand{\ie}{{\it i.e.}}
\preprint{\#HUTP-93/A029\\ 10/93}
\title{
A Bound on $m_\eta/m_{\eta'}$ for Large $N_C$\thanks{Research supported in
part by the National Science Foundation,
under grant \# PHY-9218167, and in part by the Texas National
Research Laboratory Commission under grant \# RGFY93-278B.}}
\author{
Howard Georgi \\
Lyman Laboratory of Physics \\
Harvard University \\
Cambridge, MA 02138 \\
}
\date{}
\abstract{If the number of colors is large, the ratio
$m_\eta/m_{\eta'}$ is bounded from above. The bound is not satisfied by the
observed $\eta$ and $\eta'$ masses.
}
\starttext

\section{\label{intro}Introduction}

One of the classic statements of the $U(1)$ problem in QCD was Weinberg's
observation that a chiral $U(1)$ broken only by quark masses (isospin
symmetric and small enough to apply chiral Lagrangian arguments) implies the
existence of a neutral meson state with a mass less than
$\sqrt3m_\pi$.~\cite{weinberg} 't~Hooft showed how nonperturbative effects
could solve this problem by breaking the chiral $U(1)$.~\cite{thooft} If the
chiral $U(1)$ is broken strongly, it does not make sense to regard the $\eta'$
as a Goldstone boson. However, if the number of colors is large (and it is
often speculated that 3 is large enough), the breaking of the chiral $U(1)$ is
suppressed and leading order chiral Lagrangian arguments can still be applied.
In this note, I review the well-known\footnote{See, for example,
\cite{veneziano}.} form for the pseudoscalar meson mass-squared matrix in this
limit and note the existence of an upper bound on the ratio
$m_\eta/m_{\eta'}$. The bound has the form
\begin{equation}
{m_\eta^2\over m_{\eta'}^2}<{3-\sqrt3\over3+\sqrt3}
+{3\sqrt3\over\left(3+\sqrt3\right)^2}\left({m_u+m_d\over m_s}\right)
+\cO\left(\left({m_u+m_d\over m_s}\right)^2\right)
\label{bound}
\end{equation}
with the $u$, $d$ and $s$ quark masses denoted by $m_u$, $m_d$ and $m_s$. What
is perhaps slightly amusing about this bound is that it is {\bf not
satisfied} by the observed $\eta$ and $\eta'$ masses. This is a clear (if not
very surprising) indication that higher order effects in the chiral Lagrangian
are very important for the $\eta$-$\eta'$ system.

In leading nontrivial order in large $N$ and the momentum expansion, the
chiral Lagrangian for the nonet of pseudoscalar mesons takes the following
form:
\be
\begin{array}{c}
\cL (\pi) = f^2 \Bigl\{ \frac{1}{4} \tr \left(\partial^\mu U^\dagger
\partial_\mu U \right)\\ +\frac{1}{2} \tr \left( U^\dagger \mu
M
\right) + \frac{1}{2} \tr \left( U \mu M \right) + \frac{1}{2}
m_0^2\left(\det U
+\det U^\dagger\right)\Bigr\}\,.
\end{array}
\label{cl1}
\ee
where
\be
\begin{array}{l}
U = \exp [2i \ltilde{\Pi} /f] \\
\ds \ltilde{\Pi} = \sum_{a=0}^8\pi_aT_a, \end{array} \label{cl2}
\ee
where $f$ is a constant with dimensions of mass and $M$ is the quark mass
matrix,
\begin{equation}
M=\pmatrix{m_u&0&0\cr 0&m_d&0\cr 0&0&m_s\cr}\,.
\label{m}
\end{equation}
The
$U$ field transforms linearly under $U(3) \times U(3)$:
\be
U \rightarrow U' = LU R^\dagger , \label{cl3}
\ee

In the basis $(u\bar u,\,d\bar d,\,s\bar s)$, (\ref{cl1}) gives a
mass-squared matrix of the flavor-neutral pseudo-Goldstone bosons proportional
to
\begin{equation}
\pmatrix{x+m_u&x&x\cr x&x+m_d&x\cr x&x&x+m_s\cr}
\label{pgm}
\end{equation}
where the $x$s arise from the $m_0^2$ term in (\ref{cl1}).~\cite{veneziano}
{}From (\ref{pgm}), I will derive the bound for $m_u=m_d=0$, where the algebra
is simple. I will then indicate how to derive most easily the result to next
order.

For $m_u=m_d=0$, the mass-squared matrix,
(\ref{pgm}), has one zero eigenvalue. The other eigenvalues are
\begin{equation}
{\frac {3\,x}{2}}+{\frac {m_s}{2}}\pm{\frac {\sqrt {9\,x^{2}-2\,xm_s+m_s^{2}}}
{2}}\,.
\label{eigenvalues}
\end{equation}
The ratio is
\begin{equation}
r_0(x,m_s)\equiv{\frac {{3\,x}+m_s-\sqrt {9\,x^2-2\,xm_s
+m_s^2}}{3\,x+m_s+\sqrt {9\,x^2
-2\,xm_s+m_s^2}}}\,.
\label{ratio0}
\end{equation}
This is maximized for $x=m_s/3$, which gives the first term in the result
(\ref{bound}).

The second term in (\ref{bound}) can be most easily obtained by setting
$m_u=m_d$
and computing the ratio of the two largest eigenvalues in perturbation theory.
It is easy to see that the general result to first order in $m_u/m_s$ and
$m_d/m_s$
depends only on $m_u+m_d$, thus no information is lost by setting $m_u=m_d$.
To first
order, the ratio is
\begin{equation}
r_0(x,m_s)+\frac{m_u+m_d}{m_s} r_1(x,m_s)
\label{ratio1}
\end{equation}
where
\begin{equation}
r_1(x,m_s)={\frac {6\,x^{2}+2\,m_s^{2}}{\left (3\,x+m_s+\sqrt
{9\,x^{2}-2\,xm_s+m_s^{2}}
\right )^{2}\sqrt {9\,x^{2}-2\,xm_s+m_s^{2}}}}\,.
\label{r1}
\end{equation}
Setting $x=m_s/3$ in (\ref{r1}) gives the second term in (\ref{bound}).

The first term in (\ref{bound}) gives a mass ratio bound of
\begin{equation}
{m_\eta\over m_{\eta'}}< 0.518
\label{fbound0}
\end{equation}
compared to the experimental value
\begin{equation}
{m_\eta\over m_{\eta'}} \approx 0.572\,.
\label{exp}
\end{equation}
Including the effects of the nonzero $u$ and $d$ masses brings these closer,
but not into agreement. Using generous values $m_d/m_s\approx0.06$ and
$m_u/m_d\approx0.7$ (in both cases probably erring on the side of increasing
$(m_u+m_d)/m_s$) gives
\begin{equation}
{m_\eta\over m_{\eta'}}< 0.540\,.
\label{fbound1}
\end{equation}

Two brief comments:

\begin{enumerate}
\item
Note the role of large $N$ in the difference between Weinberg's bound
(\cite{weinberg}) and (\ref{bound}). To obtain Weinberg's bound, you
maximize the ratio of $m_\eta$ to $m_\pi$ under variations of the ratio of the
decay constants of the octet and singlet pseudoscalars. In (\ref{bound}), the
ratio of decay constants is fixed to 1 by large $N$, and what varies is the
ratio of $m_s$ to the nonperturbative contribution to $m_{\eta'}$.
\item
It is not surprising that the large $N$, chiral perturbation theoretic
analysis fails for the $\eta'$. The $\eta'$ mass in our world is sufficiently
large that higher order terms in the chiral Lagrangian are probably important.
Likewise, three colors is surely not enough to justify total neglect of
nonleading terms in $1/N$. Nevertheless, I find it amusing that the failure
happens the way it does. It is not that you can fit the masses and then the
details like decay rates and branching ratios don't work. You can't even get
started.
\end{enumerate}

\section*{Acknowledgements}
I am grateful to Chris Carone, John Donoghue, Mitch Golden and Misha Voloshin
for useful discussions and suggestions.

\end{document}